\documentclass[preprint,12pt, a4paper]{elsarticle}

\usepackage{amssymb}

\usepackage{lineno}
\usepackage{hyperref}

\usepackage[utf8]{inputenc}
\usepackage{alphabeta}

\usepackage[frozencache=true]{minted} 
\newmintinline[jl]{julia}{}
\newminted[jlcode]{julia}{linenos, firstnumber=last, breaklines, xleftmargin=\parindent, frame=lines}

\usepackage{float}
\restylefloat{table}

\newlength\columnlistingwidth
\usepackage{soul}

\usepackage{color}

\journal{SoftwareX}

\begin{document}

\begin{frontmatter}



\title{PowerDynamics.jl - An experimentally validated open-source package for the dynamical analysis of power grids}


\author[label1,label2]{Anton Plietzsch}\author[label1,label2]{Raphael Kogler}\author[label3]{Sabine Auer}\author[label4]{Julia Merino}\author[label4]{Asier Gil-de-Muro}\author[label3]{Jan Liße}\author[label3]{Christina Vogel}\author[label1]{Frank Hellmann}

\address[label1]{Potsdam Institute for Climate Impact Research}
\address[label2]{Humboldt-Universität zu Berlin}
\address[label3]{elena international GmbH}
\address[label4]{TECNALIA, Basque Technology and Research Alliance (BRTA)}

\begin{abstract}

PowerDynamics.jl is a Julia package for time-domain modeling of power grids that is specifically designed for the stability analysis of systems with high shares of renewable energies. It makes use of Julia’s state-of-the-art differential equation solvers and is highly performant even for systems with a large number of components. Further, it is compatible with Julia’s machine learning libraries and allows for the utilization of these methods for dynamical optimization and parameter fitting. The package comes with a number of predefined models for synchronous machines, transmission lines and inverter systems. However, the strict open-source approach and a macro-based user-interface also allows for an easy implementation of custom-built models which makes it especially interesting for the design and testing of new control strategies for distributed generation units. This paper presents how the modeling concept, implemented component models and fault scenarios have been experimentally tested against measurements in the microgrid lab of TECNALIA.

\end{abstract}

\begin{keyword}
Power Systems \sep Dynamic Simulation \sep Microgrids \sep Inverters \sep Julia



\end{keyword}

\end{frontmatter}

\section*{Required Metadata}
\label{Metadata}


\begin{table}[H]
\begin{tabular}{|l|p{6.5cm}|p{6.5cm}|}
\hline
C1 & Current code version & v2.4.1 \\
\hline
C2 & Permanent link to code/repository used for this code version & \url{https://github.com/JuliaEnergy/PowerDynamics.jl} \\
\hline
C3 & Legal Code License   &  GPL-3.0 License  \\
\hline
C4 & Code versioning system used & git \\
\hline
C5 & Software code languages, tools, and services used & Julia\\
\hline
C6 & Compilation requirements, operating environments \& dependencies & see Project.toml \\
\hline
C7 & Link to developer documentation &  \url{https://juliaenergy.github.io/PowerDynamics.jl/stable} \\
\hline
C8 & Questions and support &
\url{https://github.com/JuliaEnergy/PowerDynamics.jl/issues}\\
\hline
\end{tabular}
\caption{Code metadata}
\end{table}




\section{Motivation and significance}
\label{sec:motivation}
The massive integration of renewable energy sources comes with challenges for the dynamic stability and control of power grids. In conventional grids the dynamical stability is ensured by controlling a relatively small number of large conventional generators that are synchronized over the high voltage transmission grid. The inertia of the rotating generator masses stabilizes the operating state against short term fluctuations. In contrast, solar and wind power plants produce much less power and are therefore typically installed in the distribution grid, interfaced by inverters with control schemes that lock onto the grid frequency. A transition towards a larger share of renewable energy sources therefore not only implies a decentralization and an increase in the number of generating units but also significantly decreases the total amount of stabilizing inertia. To tackle this issue, the development of more decentralized control schemes and so-called grid-forming inverter controls has recently become a very active research field~\cite{schiffer15_2}.\\

However, models of grid-connected inverters represent black-box models, do not allow a transparent verification of the model results and therefore do not allow a fast transfer of the latest research results. As CIGRE also emphasizes, different inverter types and manufacturer models are often available in different software and in different degrees of complexity. Therefore, harmonization in a common simulation model is impractical \cite{Cigre-Inverter-Modeling}.\\

At the same time, the energy transition requires rapid development and continuous improvement of reliable simulation tools for dynamic network analysis \cite{auer2018stability}. For this reason, Open-Source (OS) solutions are an important contribution to the implementation of a grid-stable energy transition. While in the area of static network analysis (for load flow calculations and optimal power flow problems) there are already numerous OS tools, there is little open source available in the area of dynamic modeling of frequency and voltage stability as well as the synchronization behavior of the network in the seconds and sub-seconds range. One reason for the lack of current OS solutions is the lack of a high-performance software environment.\\

PowerDynamics.jl \cite{tkittelWIW2018} is an open source software for the dynamic simulation of energy systems that aims to fill this gap \cite{AuerWIW2018}. It is based on the programming language Julia \cite{bezanson2017julia} and developed in a collaboration between the Potsdam Institute for Climate Impact Research (PIK) and the company elena international GmbH\footnote{\url{https://www.elena-international.com/}}. PowerDynamics has been shown to easily beat the simulation times of common commercial simulation environments, such as PowerFactory or MATLAB Simulink \cite{liemann2020probabilistic}. This opens up the possibility to run very large test cases or apply Monte Carlo sampling based methods \cite{menck2013basin, hellmann2016survivability}.\\

PowerDynamics comes with a library of predefined component types for generators, loads, inverters, transmission lines and transformers, as well as a number of fault scenarios. Users can easily build their own test cases or read in existing test case files. We encourage users to also implement their own component types and eventually share it with others by making a pull request on GitHub\footnote{\url{https://github.com/JuliaEnergy/PowerDynamics.jl}}.\\

The construction of the differential equation system is based on NetworkDynamics.jl \cite{lindner2020networkdynamics}, a Julia package for simulating large dynamical systems on complex network structures that is also developed at PIK. The numerical integration is based on Julia's DifferentialEquations.jl package \cite{rackauckas2017differentialequations}. This makes PowerDynamics also compatible with Julia's scientific machine learning packages\footnote{\url{https://sciml.ai/}} and thereby opens up the possibility to apply machine learning methods to transient stability analysis.

\section{Software description}
\label{sec:software_description}


\subsection{Software Architecture}
\label{sec:software_architecture}

\begin{figure}[ht]
  \centering
  \includegraphics[width=\textwidth]{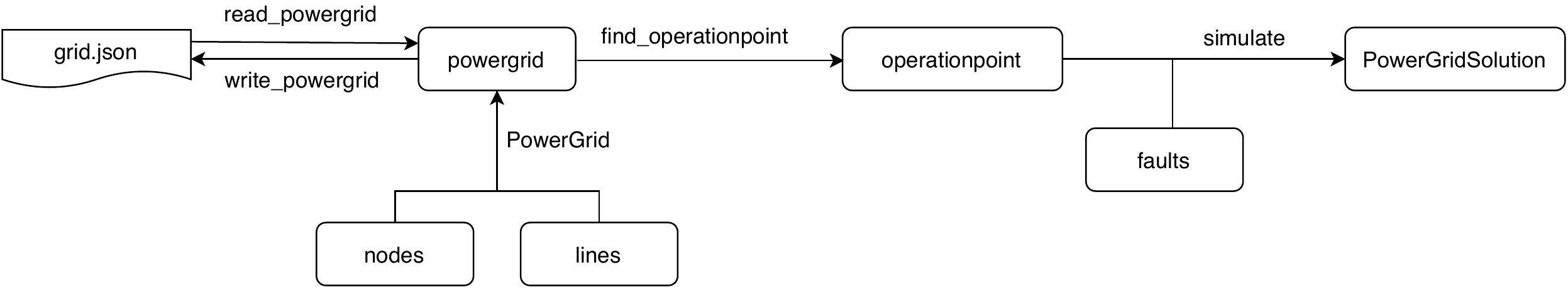}
  \caption{Simple data and workflow between components of PowerDynamics.jl}
  \label{fig:dataflow}
\end{figure}

The general workflow for a dynamic simulation with PowerDynamics.jl is shown in Fig.~\ref{fig:dataflow}. A power grid is defined by specifying the grid structure and the dynamic equations for the node and line components. Once a power grid is defined, the data can be stored in a .json-file by using \jl{write_powergrid} and loaded again by using \jl{read_powergrid}. The function \jl{find_operationpoint} is used to find fix points of the dynamical system. Different fault scenarios can be applied to the system in operation state. 

\subsubsection{Nodes}
Synchronous machines, inverters and loads are considered as \emph{node components}.
PowerDynamics.jl contains a library of already implemented node components. Additionally, users can define their own node components by using the node macro \jl{@DynamicNode}.

\subsubsection{Lines}
Transformers and electrical lines are considered as \emph{line components}. As for the node components, PowerDynamics.jl includes a standard library of different line and transformer types, including a static admittance line, \jl{StaticLine}, a dynamic admittance line, \jl{RLLine}, the (static) \jl{PiModelLine} and a simple transfomer model based on the Pi-model.

\subsubsection{Grid structure}
The \emph{PowerGrid component} is built from nodes and lines. It contains all information about the graph and is used to build the right-hand-side \jl{ODEFunction}  with the \emph{NetworkDynamics.jl}-library. This way the model definitions and the simulation engine are decoupled.

\subsubsection{Operationpoint}
The \emph{operationpoint} represents the steady-state solution of the dynamic power system. There are different methods for finding the operation point: rootfind, nlsolve and dynamic. The function \jl{find_operationpoint} returns the operationpoint which is of Type \jl{State}.

\subsubsection{Fault scenarios}
All simulation in PowerDynamics assume some kind of \emph{fault scenario}. It can be either a change in initial conditions (different from the operation point of the system) at the beginning of a simulation. Or it is a perturbation of node or line paramers for a certain time span \jl{tspan_fault} that alter the PowerGrid component. These can be e.g. a sudden change in load, \jl{PowerPerturbation}, or a line failure, \jl{LineFailure}. The latter faults are all derived from \jl{AbstractPerturbation}, and all share the same \jl{simulate} function.

\subsubsection{Numerical solution \& plotting}

The simulation of a fault scenario is done with the function \jl{simulate}. It converts the problem into an \jl{ODEProblem}, numerically integrates it using the DifferentialEquations.jl package and returns a \jl{PowerGridSolution}. This solution can then be easily plotted with Plots.jl using a predefined plotting recipe for the solution type.

\subsection{Software Functionalities}
\label{sec:software_funtionalities}


PowerDynamics.jl allows for modeling symmetric 3-phase power grids in dq-coordinates. Voltage and current are both represented by complex phasors, where the real part represents the d-coordinate and the imaginary part the q-coordinate. The advantage of this complex representation is that the complex power can be calculated by $s = p+j q = u \cdot i^*$.

The package is designed for transient stability analysis. Dynamics on shorter timescales such as harmonics and inverter switching cannot be modeled using the phasor approach.

\subsection{Sample code snippets}
\label{sec:sample_code}
The following code snippet is an explanatory example of how to use the \jl{DynamicNode}-Macro to define new inverters, generators or loads in PowerDynamics. It starts with the definition of parameters as arguments and the definition of the mass matrix. The boolean is true for differential equations and false for algebraic constraints. The assertion statements on the parameters is followed by a list of dynamics variables and finally the ODE-system itself. The ODE system is expected to contain the right-hand-side for all dynamic variables stated before. The complex voltage \jl{u} is always required for all dynamic nodes, however it can be replaced by an algebraic constraint by setting \jl{m_u=false}. The given example of a grid-informing inverter, \jl{VSIVoltagePT1}, corresponds to the one visualized in the block diagram in Fig.\ref{fig:gfi_blockdiagram}.\\

\begin{listing}[ht]
\begin{jlcode}
@DynamicNode VSIVoltagePT1(τ_v,τ_P,τ_Q,K_P,K_Q,V_r,P,Q)  begin
    MassMatrix(m_u = true, m_int = [true,true])
end
begin
    @assert τ_v > 0 "time constant voltage droop delay should be >0"
    @assert τ_P > 0 "time constant active power measurement should be >0"
    @assert τ_Q > 0 "time constant reactive power measurement should be >0"
    @assert K_Q > 0 "reactive power droop constant should be >0"
    @assert K_P > 0 "active power droop constant reactive power measurement should be >0"
end [[ω, dω],[q_m,dq_m]] begin
    p = real(u * conj(i))
    q = imag(u * conj(i))
    dφ = ω
    v = abs(u)
    dv = 1/τ_v*(-v+V_r - K_Q*(q_m-Q))
    dq_m = 1/τ_Q*(q-q_m)
    du = u * 1im * dφ + dv*(u/v)
    dω = 1/τ_P*(-ω-K_P*(p-P))
end
\end{jlcode}
\caption{Source code for a droop controlled voltage source inverter as described by Schiffer et al. \cite{schiffer13_2}}
\end{listing}

\subsection{Future Developments \& Improvements}
\begin{figure}[ht]
    \centering
    \includegraphics[width=0.9\textwidth]{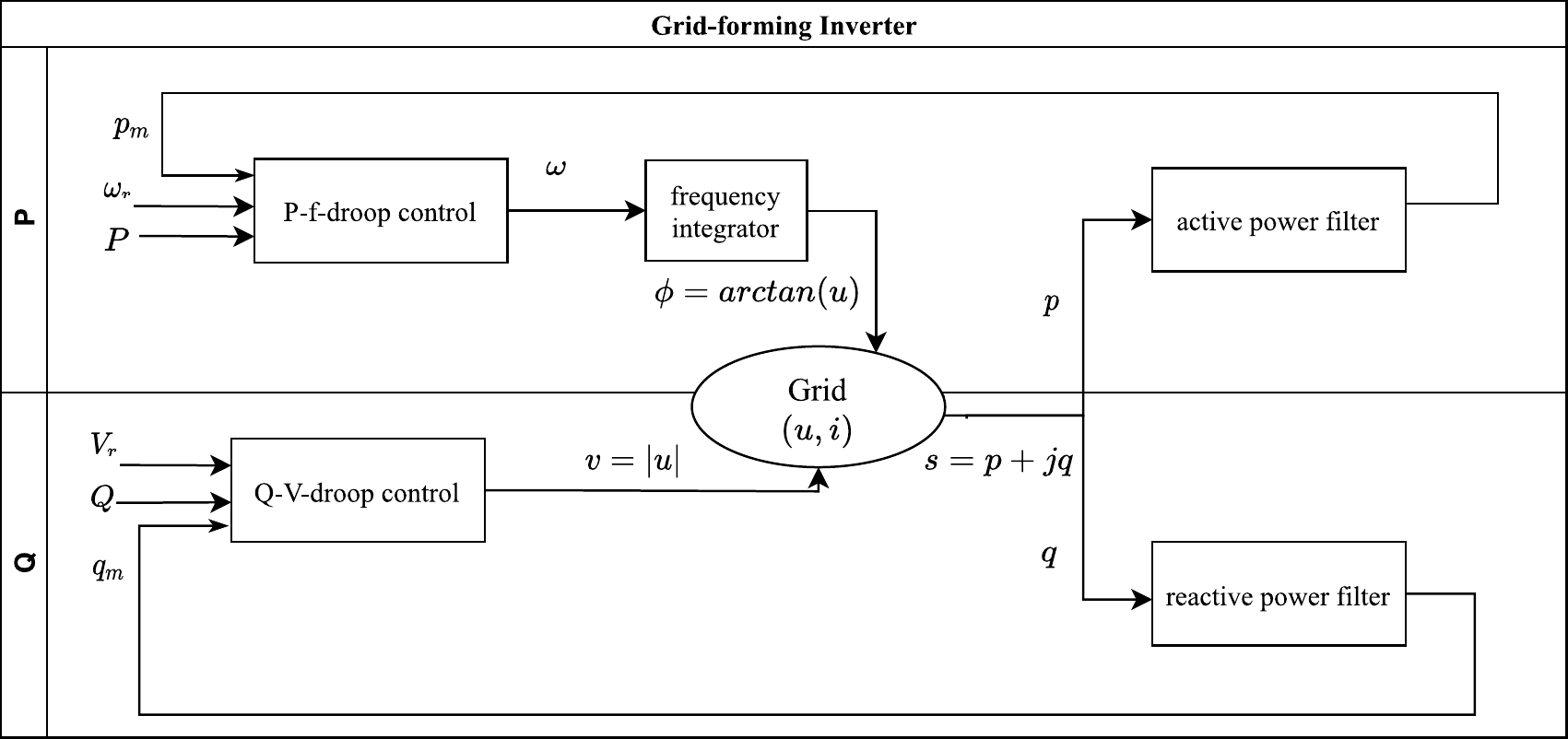}
    \caption{Example block diagram for a grid-forming inverter.}
    \label{fig:gfi_blockdiagram}
\end{figure}

The PowerDynamics community has just started to work heavily on the modularization of PowerDynamics such that power generators, inverters and other equipment can be represented in the well-known block diagram structure. We plan to undertake this modularization with the help of ModelingToolkit.jl\footnote{\url{https://github.com/SciML/ModelingToolkit.jl}}. In Figure \ref{fig:gfi_blockdiagram} a block diagram is shown for the example of a grid-forming inverter. It consists of active- and reactive power droop control (as an outer control loop), a frequency integrator and active as well as a reactive power filters. The causal modeling representation from the block diagram is planned for the new major PowerDynamics release.

Also, work in progress (with an open pull request on GitHub) is the integration of PowerModels \cite{coffrin2018powermodels}. There we assign each dynamic component a static component in order to enable an operation point search as a solution of a load flow analysis. This work is done in collaboration with TU Delft.\\

We further plan to implement more fault scenarios, such as noise perturbations \cite{auer2017stability,plietzsch2019generalized} and dynamically induced cascading failures \cite{schafer2018dynamically}. These enhancements will make use of the state-of-the-art stochastic differential equations solvers and callback functions of DifferentialEquations \cite{rackauckas2017differentialequations}. 

We also want to extend the library of component models for lines and nodes. Here, we also expect user contributions for various inverter controls that can be shared via pull requests on GitHub.\footnote{\url{https://github.com/JuliaEnergy/PowerDynamics.jl/pulls}}

More future contributions are expected from the MARiE-project (Dynamic modeling for analysis and control of intelligent energy networks) undertaken by BTU Cottbus together with elena international GmbH and funded by the German Federal Ministry of Economics and Technology.
The key innovation of the project is the development of canonical base models of grid-connected power converters by BTU Cottbus. Together with elena international, the models are then implemented in Julia and integrated in PowerDynamics.jl. To ensure practical relevance of the models they are finally  experimentally validated with a Power-in-the-loop test bed by BTU Cottbus.

\section{Illustrative Example}
\label{sec:example}



Our illustrative example is the simulation of a tripping line in the IEEE 14-bus system. This system contains 5 generators, 11 loads, 17 lines and 3 transformers. We model the loads as constant power loads and the generators by a 4th order generator model \cite{sauer1998power}. The parameters are taken from \cite{kodsi2003modeling}.

The implementation of this introductory example can be found in the packages PowerDynamicsExample\footnote{https://github.com/JuliaEnergy/PowerDynamicsExamples}. It is available both as a Julia script and a Jupyter Notebook. The latter can also be directly launched in the browser by using BinderHub.

\begin{figure}[ht]
\centering
\includegraphics[width=1.0\columnwidth]{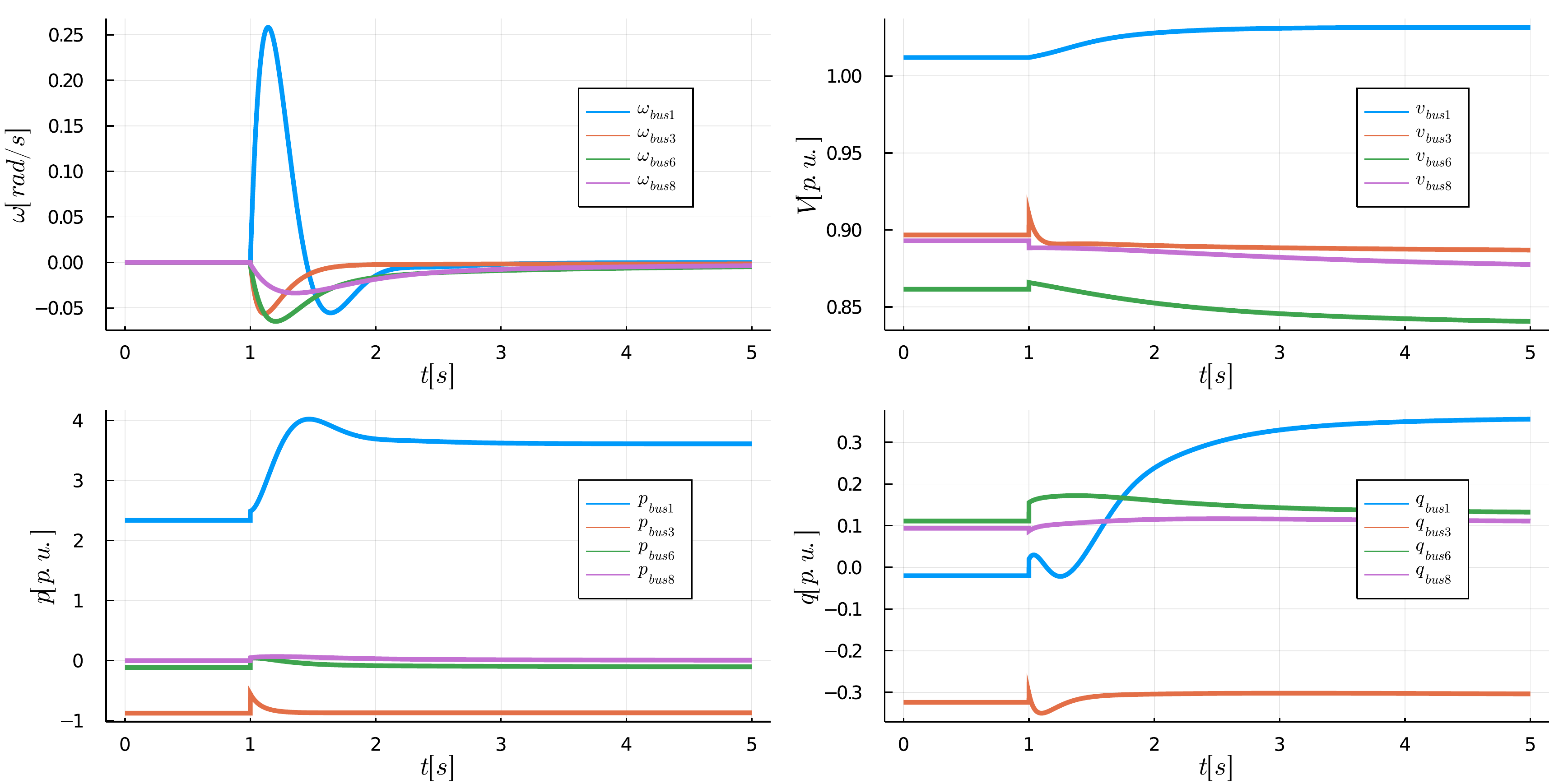}
\caption{\label{fig:ieee14} \textbf{Simulation of a line tripping in the IEEE 14-bus system.}}
\end{figure}

The simulation of this test case is straightforward and requires only a few lines of code. PowerDynamics is a registered package. This means it can be directly installed from the Julia REPL with:
\begin{listing}
\begin{jlcode}
using Pkg
Pkg.add("PowerDynamics")
\end{jlcode}
\caption{Installing PowerDynamics}
\end{listing}
For creating a new test case we would have to define an \jl{Array} or \jl{OrderedDict} of buses and lines and execute the function \jl{PowerGrid(buses,lines)}. Here, we assume that the IEEE-14bus testcase is already saved as a Json file that we can read in with \jl{read_powergrid}. As the initial state for the simulation we determine the operation point with \jl{find_operationpoint}. The fault scenario is the predefined function \jl{LineFailure}, which removes a line from the grid. The numerical integration is done with the \jl{simulate} function and the results are shown in Fig.~\ref{fig:ieee14}.

\begin{listing}
\begin{jlcode}
using PowerDynamics
powergrid = read_powergrid("ieee14-4th-order.json", Json)
operationpoint = find_operationpoint(powergrid)
timespan= (0.0,5.)
fault = LineFailure(line_name="branch2", tspan_fault=(1.,5.))
solution = simulate(fault, powergrid, operationpoint, timespan)
include("plotting.jl")
plot = create_plot(solution)
display(plot)
\end{jlcode}
\caption{Source code for simulating the IEEE14 grid with PowerDynamics}
\end{listing}

\section{Software Validation}

\begin{figure}[ht]
    \centering
    \includegraphics[width=0.8\textwidth]{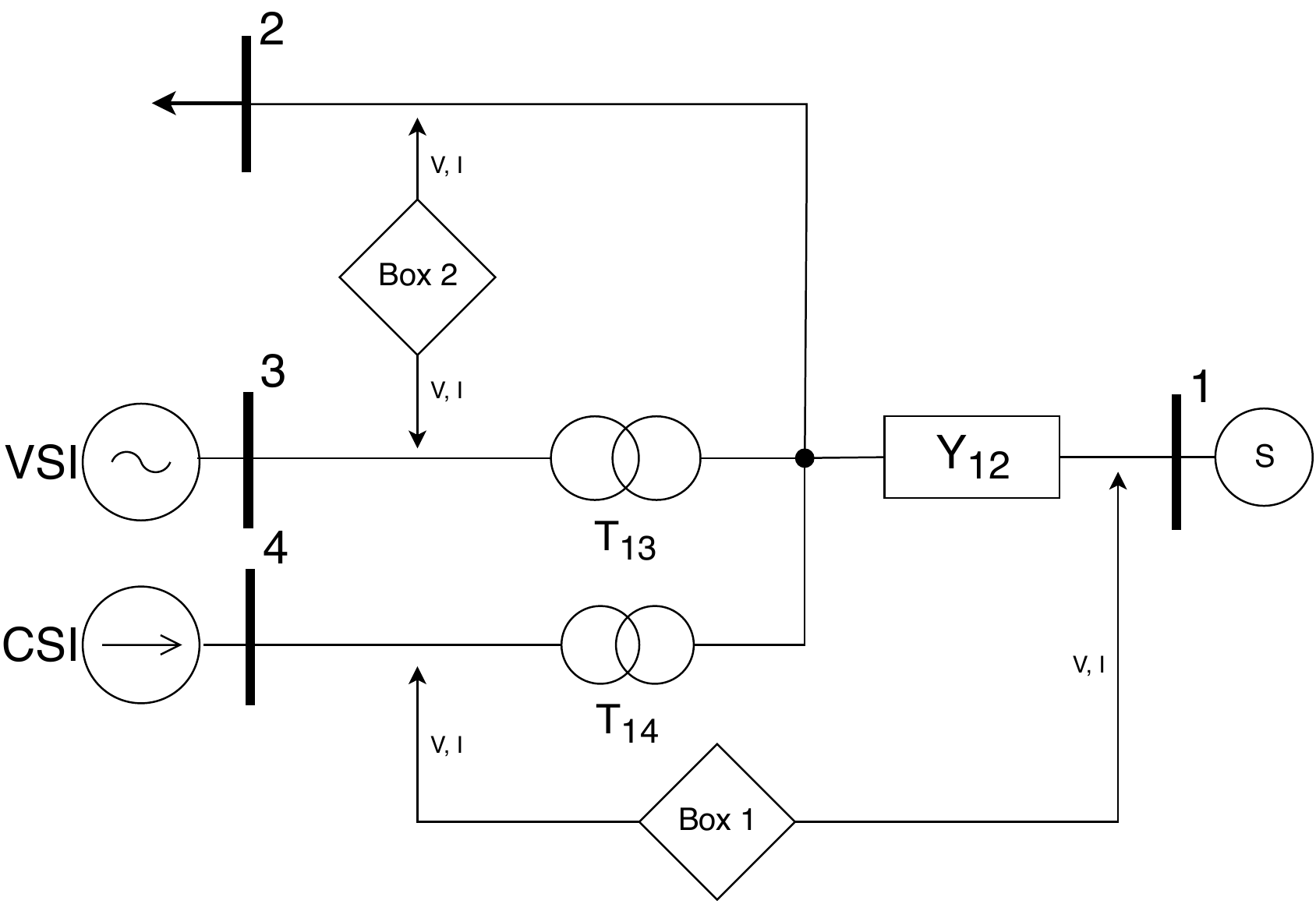}
    \caption{\textbf{Test grid setup in the laboratory.} The setup includes a grid-forming voltage source inverter (bus 3) and a grid-following current source inverter (bus 4) that are connected to the grid by transformers, a load (bus 2) and a line with admittance $Y_{12}$. The test grid has an interconnection to the Spanish grid (bus 1) which we model as a slack node. Measurement are taken out with two Boxes that can measure voltage and current signals for two phases. We assume the phases to be balanced and thereby calculate the third phase.}
    \label{fig:testcases_diagrams}
\end{figure}

\subsection{Testlab description}
An experimental validation of PowerDynamics.jl was undertaken in January 2020 at the Smart Grid Technologies Laboratory at TECNALIA within the context of the ERIgrid\footnote{\url{https://erigrid.eu/}}-funded Valeria project. The primary goal was to implement detailed dynamical models for different inverter controls, loads and perturbation scenarios in PowerDynamics.jl and to validate the numerical simulations against experimental measurements of different dynamical scenarios in a small test grid setup.\\

The laboratory at TECNALIA comes with a grid-forming and a grid-following inverter that both have self-built control schemes \cite{planas2012stability, planas2013design}. Therefore, the dynamic equations and parameters are known and can be translated into PowerDynamics code. The grid-following control consists of a low-pass filter for the voltage signal, a phase-locked loop (PLL) and a droop control for active and reactive power. The grid-forming control consists of several filters for the voltage and current signals, a droop control for frequency and voltage amplitude and a fictitious impedance.

\subsection{Test Case description}

\begin{figure}[htpb]
    \centering
    \includegraphics[width=\textwidth]{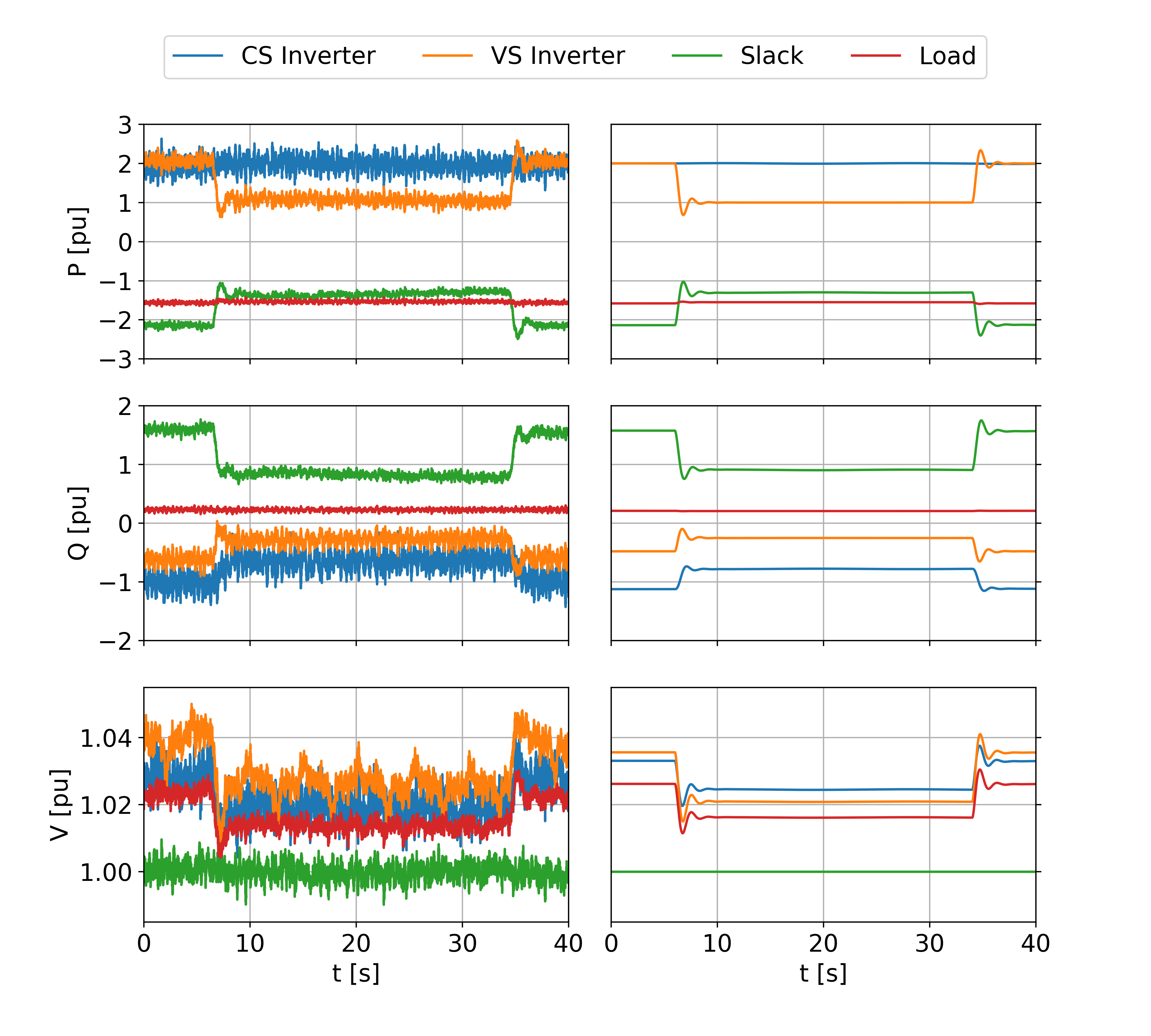}
    \caption{\textbf{Comparison of measurements and simulations.} The scenario is a set point dispatch for the active power at the grid-forming inverter in the test grid setup shown in Fig.~\ref{fig:testcases_diagrams}. Voltage and power signals are calculated from the measured current and voltage signals. The dynamic models for grid-forming and grid-following inverter have been implemented and simulated in PowerDynamics.jl.}
    \label{fig:meas_vs_sim}
\end{figure}

For demonstrative purposes in this paper we chose a test case setup including a grid-forming, a grid-following and a load in grid connected mode (Fig.~\ref{fig:testcases_diagrams}). The comparison between measurements and simulations for the scenario of a power set point dispatch at the voltage source inverter is shown in Fig.~\ref{fig:meas_vs_sim}. A more detailed description and a larger variety of test cases can be found in the technical report of the Valeria project \cite{VALERIA}.

\section{Impact}
\label{sec:impact}


\subsection{Scientific impact}


Programming languages, such as Python or the commercial language MATLAB, are not up to the special requirements of modeling renewable power grids. This includes the integration of stochastic differential equations for the representation of fluctuating renewables \cite{auer2018stability} as well as the simulation of delayed differential equations to represent the influence of measurement and delay times of inverter controls \cite{schiffer16_2}. Due to Julia's just-in-time compilation, PowerDynamics has also strong performance advantages compared to simulations in MATLAB or Python. It has also been shown to be much more performant than its commercial competitors MATLAB Simulink and DIgSILENT PowerFactory \cite{liemann2020probabilistic}. This does not only enable the dynamic simulation of very large test cases with a large number of nodes but also Monte Carlo sampling based methods. This includes for example efficient calculations of stability and performance measures such basin stability \cite{menck2013basin} and survivability \cite{hellmann2016survivability} that complement standard linear stability analysis and capture also the nonlinear behavior of the  system. Additionaly, PowerDynamics is also compatible with Julia's scientific machine learning packages such as DiffEqFlux.jl \cite{rackauckas2019diffeqflux}, a package for combining differential equations with neural networks. This will eventually lay the foundation for a new branch of research, a machine learning based energy system stability analysis \cite{nauck2020thesis}.\\


While in the context of static power flow analysis there are already numerous OS tools (which are mostly Python-based), e.g. PyPSA \cite{PyPSA} and panda-power \cite{thurner2018pandapower}, there is little open source available in the area of dynamic modeling of frequency and voltage stability as well as the synchronization behavior of the network in the seconds and subseconds range. Only the PSAT library which has not been maintained for 10 years and therefore does not contain any current inverter models, and the new library PowerSimulationDynamics.jl\footnote{\url{https://github.com/NREL-SIIP/PowerSimulationsDynamics.jl}} (another Julia package, continuation of LITS.jl \cite{henriquez2020lits}) are known. PowerSimulationDynamics, developed by NREL, uses the parsing capabilities of PowerSystems to build their industry standard power system data structures. The package is using implicit and explicit DAE representations in combination with the Sundials solvers \cite{hindmarsh2005sundials}. 
In contrast PowerDynamics is based on the DifferentionalEquations package which enables to simulate SDEs, DDEs and explicit DAEs in mass matrix representation. This allows the modeling of inverter measurement delays and renewable power fluctuations.
Additionally, since PowerDynamics uses the NetworkDynamics package for simulation model building which decouples the representation of nodes and lines, major performance optimizations with parallelization and GPU calculations are possible. This gives the possibility to model large-scale networks.
Currently, we are in contact with the developers of PowerSimulationDynamics and collaborate on the integration of ModelingToolkit\footnote{\url{https://github.com/SciML/ModelingToolkit.jl}} and further topics.


PowerDynamics is still very young and under heavy development but the user base is steadily growing. At the moment the software is developed by and used for research at PIK, at TU Berlin, at BTU Cottbus, at TU Delft and at FZ Jülich. Further, TU Delft plans to use PowerDynamics for teaching electrical engineering courses. \\

\subsection{Commercial impact}

PowerDynamics.jl will serve as a basis or backend for web applications and other commercial products of elena international which is a spin-off of the Potsdam-Institute of Climate Impact Research. In particular, PowerDynamics together with existing Open-Source tools for static network analyses will be the backend of an innovative web application which will be offered as an analysis tool for the renewable conversion of power systems first in islanded and then also in interconnected operation. Here elena int. is already working on a prototype for the planning of micro power grids. This software can then be used for the planning of micro power grids (in the global south) and for the new concept of cellular energy systems in Europe.

As an Open-Source Software PowerDynamics.jl was also already used for creating an online simulation environment for an online course about ``inertia requirements for renewable power systems''\footnote{\url{https://www.renac.de/trainings-services/trainings/ready-made-trainings/product/inertia-requirements-for-renewable-power-systems}}. The goal of the course is to guide participants through the topic of inertia issues and possible solutions for highly renewable power grids\footnote{\url{https://github.com/SabineAuer/OnlineCourse-Inertia}}.

\section{Conclusions}
\label{sec:conclusions}
PowerDynamics is an OS Julia package for transient stability analysis which is both suitable for highly renewable power grids, especially relevant for microgrids, and large-scale networks with a large number of buses due to its performance advantages.
It allows for the integration of delays and fluctuation dynamics of renewable power sources. Due to its decoupled representation of node and line components, it can be parallelized and allows for future performance optimization with GPU calculations.
PowerDynamics is still a young library, however, several research projects have started that relate to PowerDynamics and push its development.
Currently, it is work in progress to integrate PowerModels for the operation point search based on a load-flow analysis. Further, work started on integrating the causal modeling approach in PowerDynamics that allows the representation of inverters as block diagrams as it is coming in control theory to bridge the gap to this community and engage a wider user base into the development and spreading of PowerDynamics.


\section{Conflict of Interest}


We wish to confirm that there are no known conflicts of interest associated with this publication and there has been no significant financial support for this work that could have influenced its outcome.

\section*{Acknowledgements}
This research has been performed using the ERIGrid Research Infrastructure and is part of a project that has received funding from the European Union’s Horizon 2020 Research and Innovation Programme under the Grant Agreement No. 654113. The support of the European Research Infrastructure ERIGrid and its partner TECNALIA is very much appreciated. We further acknowledge the support by BMBF (CoNDyNet2 FK. 03EK3055A), the DFG (ExSyCo-Grid, 410409736) and the Leibnitz competition (T42/2018).





\bibliographystyle{elsarticle-num}
\bibliography{references}







\end{document}